\documentclass[proceedings]{JHEP3}

\conference{International Europhysics Conference on HEP}        

\title{Exclusive and Semi-inclusive $B$ Decays in QCD Factorization}

\author{{Hai-Yang Cheng}\\               
        Institute of Physics, Academia Sinica, Taipei, Taiwan 115, R.O.C.\\        
        E-mail: \email{phcheng@ccvax.sinica.edu.tw}}               

\abstract{Applications of QCD factorization to $B\to\phi K$,
charmless $B\to VV$, $B\to \J K(K^*)$ and semi-inclusive decays
$B\to MX$ are discussed.}

\def\be{\begin{eqnarray}}
\def\en{\end{eqnarray}}
\def\non{\nonumber}
\def\la{\langle}
\def\ra{\rangle}

\def\J{{J/\psi}}
\def\ov{\overline}

\begin{document}

  \section{$B\to\phi K$ decay}
Recently CLEO \cite{Briere}, Belle \cite{Bozek} and BaBar
\cite{BaBar} have reported the results:
 \be
 {\cal B}(B^\pm\to\phi K^\pm)=\cases{
(5.5^{+2.1}_{-1.8}\pm0.6)\times 10^{-6} & CLEO, \cr
(7.7^{+1.6}_{-1.4}\pm0.8)\times 10^{-6} & BaBar, \cr
(11.2^{+2.2}_{-2.0}\pm1.4)\times 10^{-6} & Belle,} \label{exp1}
 \en
and
 \be
 {\cal B}(B^0\to \phi K^0)=\cases{ <12.3\times 10^{-6} & CLEO, \cr
 (8.1^{+3.1}_{-2.5}\pm0.8)\times
10^{-6} & BaBar, \cr (8.9^{+3.4}_{-2.7}\pm1.0)\times 10^{-6} &
Belle.} \label{exp2}
 \en
The neutral mode $B^0\to\phi K^0$ is a pure penguin process, while
the charged mode $\phi K^-$ receives an additional annihilation
contribution which is quark-mixing-angle suppressed. The predicted
branching ratio is very sensitive to the nonfactorizable effects
which are sometimes parameterized in terms of the effective number
of colors $N_c^{\rm eff}$; it falls into a broad range $(13\sim
0.4)\times 10^{-6}$ for $N_c^{\rm eff}=2\sim\infty$ \cite{CY99}.
Therefore, a theory calculation of the nonfactorizable corrections
is urgently needed in order to have a reliable prediction which
can be used to compare with experiment.

In QCD factorization approach, the branching ratio is predicted to
be ${\cal B}(B^\pm \to\phi K^\pm)=(4.0\pm 0.8)\times 10^{-6}$ in
the absence of annihilation contributions, where theoretical error
comes from the logarithmic divergent term occurring in spectator
interactions
$$X_H\equiv\int^1_0 {dx\over x}=\ln{M_B\over \Lambda_{\rm QCD}}
(1+\rho_H),~~~\rho_H\leq 1$$ Power-suppressed annihilation is
often treated to be negligible based on helicity suppression
argument. However, annihilation diagrams induced by $(S-P)(S+P)$
penguin operators are not subject to helicity suppression.
Including theoretical errors from both $X_H$ for spectator
interactions and $X_A$ for weak annihilations, we obtain
\cite{CYphiK}
 \be
 {\cal B}(B^-\to\phi K^-)=\,(4.6^{+3.2}_{-1.5})\times
 10^{-6},\qquad {\cal B}(B^0\to\phi K^0)=\,(4.2^{+3.0}_{-1.3})\times
 10^{-6}.
 \en
Hence, the prediction is in agreement with data within
experimental and theoretical errors.

Recently, calculations within the framework of pQCD are also
available \cite{pqcd}. The pQCD results ${\cal B}(B^-\to\phi
K^-)=\,(10.2^{+3.9}_{-2.1})\times 10^{-6}$ and ${\cal
B}(B^0\to\phi K^0)=\,(9.6^{+3.7}_{-2.0})\times 10^{-6}$ are large
for two reasons. First, the relevant scale in the pQCD calculation
is $\mu\sim 1.5$ GeV and the relevant Wilson coefficient
$c_4(\mu)$ at this low scale increases dramatically as $\mu$
decreases. However, such a ``dynamic enhancement" does not exist
in QCD factorization because the parameter $a_4$ which contains
the term $c_4(\mu)+c_3(\mu)/3$ is formally renormalization scale
and $\gamma_5$ scheme independent after including ${\cal
O}(\alpha_s)$ vertex-type and penguin-type corrections. Second,
the contribution from the chromomagnetic dipole operator to $a_4$,
which is absent in the current pQCD calculation, is sizable but
destructive.  Therefore, a refined measurement of $B\to \phi K$
decays will provide a nice ground for discriminating between the
approaches of QCD factorization and pQCD.

\section{Charmless $B\to VV$ decays}
It is known that the decay amplitude of a $B$ meson into two
vector mesons  is governed by three unknown form factors
$A_1(q^2),~A_2(q^2)$ and $V(q^2)$ in the factorization approach.
It has been pointed out in \cite{CCTY} that the charmless $B\to
VV$ rates are very sensitive to the form-factor ratio $A_2/A_1$.
This form-factor ratio is almost equal to unity in the
Bauer-Stech-Wirbel (BSW) model \cite{BSW}, but it is less than
unity in the light-cone sum rule (LCSR) analysis for form factors
\cite{Ball}. In general, the branching ratios of $B\to VV$
predicted by the LCSR are always larger than that by the BSW model
by a factor of $1.6\sim 2$ \cite{CCTY}. This is understandable
because in the heavy quark limit, both vector mesons in the
charmless $B\to VV$ decay should have zero helicity and the
corresponding amplitude is proportional to the form factor
difference $(A_1-A_2)$. These two form factors are identical at
$q^2=0$ in the BSW model.

\begin{table}[ht]
\begin{center}
\begin{tabular}{|l |l c l |} \hline
 Decay & LCSR & BSW & Expt.
\\ \hline
  $ B^- \to K^{*-} \phi$
 & 9.30 & 4.32 & $9.7^{+4.2}_{-3.4}\pm1.7$ (BaBar) \\
  & & & $10.6^{+6.4+1.8}_{-4.9-1.6}$ (CLEO) \\
  & & & $<18$ (Belle) \\
  \hline
   $ \ov B^0 \to\ov K^{*0} \phi$
  & 8.71 & 4.62 & $8.6^{+2.8}_{-2.4}\pm1.1$ (BaBar)\\
  & & & $11.5^{+4.5+1.8}_{-3.7-1.7}$ (CLEO) \\
  & & & $13.0^{+6.4}_{-5.2}\pm 2.1$ (Belle) \\
  \hline
\end{tabular}
\end{center}
\caption{Branching ratios (in units of $10^{-6}$) for $B\to
K^*\phi$ modes. Two different form-factor models, the LCSR and the
BSW models, are adopted and the unitarity angle $\gamma=60^\circ$
is employed. Experimental results are taken from
\cite{Briere,Bozek,BaBar}.}
\end{table}

We have analyzed $B\to VV$ decays within the framework of QCD
factorization \cite{CYvv}. We  see from Table I that the first
observed charmless $B\to VV$ mode, $B\to\phi K^*$, recently
measured by CLEO \cite{Briere}, Belle \cite{Bozek} and BaBar
\cite{BaBar}, clearly favors the LCSR over the BSW model for $B-V$
transition form factors. Contrary to phenomenological generalized
factorization, nonfactorizable corrections to each partial-wave or
helicity amplitude are not the same; the effective parameters
$a_i$ vary for different helicity amplitudes. The leading-twist
nonfactorizable corrections to the transversely polarized
amplitudes vanish in the chiral limit and hence it is necessary to
take into account twist-3 distribution amplitudes of the vector
meson in order to have renormalization scale and scheme
independent predictions. Owing to the absence of $(S-P)(S+P)$
penguin operator contributions to $W$-emission amplitudes,
tree-dominated $B\to VV$ decays tend to have larger branching
ratios than the penguin-dominated ones \cite{CYvv}. For example,
$B^0\to \rho^+\rho^-$ has a branching ratio of order $4\times
10^{-5}$.

\section{$B\to\J K(K^*)$ decays}
The hadronic decays $B\to \J K(K^*)$ are interesting because
experimentally they are a few of the color-suppressed modes which
have been measured, and theoretically they are calculable by QCD
factorization even the emitted meson $\J$ is heavy. That is, this
is the only color-suppressed mode that one can compute and compare
with experiment.

To leading-twist contributions from the light-cone distribution
amplitudes (LCDAs) of the mesons, vertex corrections and hard
spectator interactions including $m_c$ effects imply $|a_2(\J
K)|\sim 0.11$ vs. 0.25 by experiment \cite{CYJpsiK}. Hence, the
predicted branching ratio is too small by a factor of 5 ; the
nonfactorizable corrections to naive factorization to
leading-twist order are small. We study the twist-3 effects due to
the kaon and find that the coefficient $a_2(\J K)$ is largely
enhanced by the nonfactorizable spectator interactions arising
from the twist-3 kaon LCDA $\phi^K_\sigma$, which are formally
power-suppressed but chirally, logarithmically and kinematically
enhanced. Therefore, factorization breaks down at twist-3 order.
It is found in \cite{CYJpsiK} that $a_2(\J
K)=0.19^{+0.14}_{-0.12}$ for $|\rho_H|\leq 1$ and that twist-2 as
well as twist-3 hard spectator interactions are equally important.

Recently, the spin amplitudes $A_0$, $A_\|$ and $A_\bot$ for $B\to
\J K^*$ decays in the transversity basis and their relative phases
have been measured by Belle \cite{BelleJ} and BaBar \cite{BaBarJ}.
The decay $B\to\J K^*$ is currently analyzed within the framework
of QCD factorization \cite{CKY} and it is found that the effective
parameters $a_2^h$ for helicity $h=0,+,-$ states receive different
nonfactorizable contributions. Contrary to the $\J K$ case,
$a_2^0$ in $B\to\J K^*$ does not receive twist-3 contributions and
it is dominated by twist-2 hard spectator interactions.

\section{Semi-inclusive $B$ decays}
The semi-inclusive decays $B\to M+X$ that are of special interest
originate from the quark level decay, $b \rightarrow M + q$. They
are theoretically cleaner compared to exclusive decays and have
distinctive experimental signatures \cite{Soni,Browder}. The
theoretical advantages are : (i) A very important theoretical
simplification occurs in the semi-inclusive decays over the
exclusive decays if we focus on final states such that $M$ does
not contain the spectator quark of the decaying $B(B_s)$ meson as
then we completely by-pass the need for the transition form factor
for $B(B_s) \to M$. (ii) There is no troublesome infrared
divergent problem occurred at endpoints when working in QCD
factorization, contrary to the exclusive decays where endpoint
infrared divergences usually occur at twist-3 level, and (iii) As
for $CP$ violation, contrary to the exclusive hadronic decays, it
is not plagued by the unknown soft phases. Consequently, the
predictions of the branching ratios and partial rate asymmetries
for $B\to MX$ are considerably clean and reliable. Since these
semi-inclusive decays also tend to have appreciably larger
branching ratios compared to their exclusive counterparts, they
may therefore be better suited for extracting CKM-angles and for
testing the Standard Model.

In order to have a reliable study of semi-inclusive decays both
theoretically and experimentally, we will impose two cuts. First,
a momentum cutoff imposed on the emitted light meson $M$, say
$p_M>2.1$ GeV, is necessary in order to reduce contamination from
the unwanted background and ensure the relevance of the two-body
quark decay $b\to Mq$. Second, it is required that the meson $M$
does not contain the spectator quark in the initial $B$ meson and
hence there us no $B-M$ transition form factors. Under these two
cuts, we argue that the factorization formula for exclusive decays
can be generalized to the semi-inclusive decay:
 \be \label{qcdfsemi}
\la MX|O|B\ra = \underbrace{\int^1_0 du\,
T^I(u)\Phi_{M}(u)}&+\underbrace{\int^1_0 d\xi \,du
\,T^{II}(\xi,u)\Phi_B(\xi)\Phi_{M}(u)}.  \non \\
{\rm no~form~factors} & {{\rm
~~~~~~power~suppressed~by}~(\Lambda_{\rm QCD} /m_b)^{1/2}}
 \en
However, this factorization formula is not as rigorous as the one
for the exclusive case. To the order ${\cal O}(\alpha_s)$, there
are two additional contributions besides vertex corrections: the
bremsstrahlung process $b\to Mq\,g$ ($g$ being a real gluon) and
the process $b\to Mq\,g^*\to Mqq'\bar q'$. The bremsstrahlung
subprocess could potentially suffer from the infrared divergence.
However, the vertex diagram in which a virtual gluon is attached
to $b$ and $q$ quarks is also infrared divergent. This together
with the above-mentioned bremsstrahlung process will lead to a
finite and well-defined correction.  This finite correction is
expected to be small as it is suppressed by a factor of
$\alpha_s/\pi\approx 7\%$. In the presence of bremsstrahlung and
the fragmentation of the quark-antiquark pair from the gluon, the
factorizable configurations $\la X_1M|j_1|0\ra\la X_1'|j_2|B\ra$
and $\la X_2|j_1|0\ra\la X_2'M|j_2|B\ra$ with $X_1+X_1'=X$ and
$X_2+X_2'=X$ are allowed. In general, one may argue that these
configurations are suppressed since the momentum cut $p_M>2.1$ GeV
favors the two-body quark decay $b\to Mq$ and low multiplicity for
$X$. However, it is not clear to us how rigorous this argument is.
Therefore, we will confine ourselves to vertex-type and
penguin-type corrections as well as hard spectator interactions so
that the factorization formula (\ref{qcdfsemi}) is applicable to
semi-inclusive decays at least as an approximation.

Some highlights of the present analysis are \cite{Bincl}:
\begin{itemize}
\item Though phase-space and power suppressed, hard spectator interactions are extremely
important for color-suppressed modes, e.g.
$(\pi^0,\rho^0,\omega)X_{\bar s},~\phi X$ and $\J X_s,\J X$. This
is because the relevant hard spectator correction is color
allowed, whereas $b\to Mq$ for these modes are color-suppressed.
\item The prediction ${\cal B}(B\to \J X_s)=9.6\times 10^{-3}$ is in
agreement with experiments: $(8.0\pm 0.8)\times 10^{-3}$ from CLEO
and $(7.89\pm0.10\pm0.40)\times 10^{-3}$ from BaBar.
\item $\ov B^0_s\to
(\pi^0,\rho^0,\omega)X_{\bar s}$, $\rho^0X_{s\bar s}$, $\ov
B^0\to(K^-X,K^{*-}X)$ and $B^-\to (K^0X_s,K^{*0}X_s)$ are the most
promising ones in searching for direct CP violation: they have
branching ratios of order $10^{-6}-10^{-4}$ and CP rate
asymmetries of order $(10-40)\%$.  With $1\times 10^7$ $B\ov B$
pairs, the asymmetry in $K^{*-}$ channel starts to become
accessible. With about $7\times 10^7$ $B\ov B$ events, the PRA's
in other modes mentioned above will become feasible.
\end{itemize}

\acknowledgments I wish to thank Kwei-Chou Yang and Amarjit Soni
for collaborations and Physics Department, Brookhaven National
Laboratory for its hospitality.



\end{document}